\documentstyle[preprint,aps,prl]{revtex}
\begin{document}
%\Artigo corrigido em 22/10/2002.
\title{Electron states in boron nitride nanocones}

\author{S\'ergio Azevedo$^{a}$, M\'ario S. C. Mazzoni$^{b}$, 
H. Chacham, R. W. Nunes }

\address
{Departamento de F\'{\i}sica, ICEX, Universidade Federal de Minas
Gerais, CP 702, 30123-970,Belo Horizonte, MG, Brazil.}

\footnotetext{$^a$On leave from 
 Departamento de F\'\i sica, 
Universidade Estadual de
Feira de Santana, Km 3 BR-116, 44031-460, Feira de Santana, BA, Brazil.}
\footnotetext{$^b$corresponding author: mazzoni@fisica.ufmg.br}

\maketitle
\begin{abstract}
We apply first-principles calculations to study the electronic
structure of boron nitride nanocones with disclinations of different angles
$\theta=n\pi/3$. Nanocones with odd values of $n$ present antiphase
boundaries  
%hc characterized by a line defect of ``wrong'' B-B or N-N bonds. These
%hc line defects lead to 
that cause
a reduction of the work function of the
nanocones, relative to the bulk BN value, by as much as 2 eV. 
%hc In contrast, nanocones with even values of $n$ do not have line
In contrast, nanocones with even values of $n$ do not have such
defects and present work functions that are very similar to the BN bulk
value. 
%hc These results may help explain the observed field emission
%hc properties of boron nitride nanocones and nanotubes.
These results should have strong consequences for the field emission
properties of boron nitride nanocones and nanotubes.

\end{abstract}

\vspace{1cm}
\noindent
pacs {71.20.Tx, 71.15.Mb, 71.24.+q}
\newpage

%{\bf I. Introduction}

Curved nanoscale structures, of which the best known examples are
carbon fullerenes and nanotubes, have been the focus of increased
scientific and technological interest, due to their unique electronic
and mechanical properties. \cite{kroto,iijima1,hamada}.  
%hc and mechanical properties. \cite{kroto,iijima1}.  
%hc The deviation from
%hc the planar geometry may have a strong effect on the electronic
%hc properties of the system. For instance, 
%hc a carbon nanotube can behave either as a
%hc semiconductor or as a metal, depending on the tube diameter and
%hc chirality \cite{hamada}. 
The incorporation of pentagonal atomic rings
and other topological defects into the hexagonal network of carbon
nanotubes increases the local curvature and can lead to the
closure of the tubes \cite{iijima2}.  The structure of the cap depends
on the specific defect included, but generally it has the aspect of a
conical surface with electronic properties that are distinct from the bulk
material \cite{charlier}.  The same applies to boron nitride (BN)
nanocones, which can be formed either as the closing caps at the edges of BN
nanotubes \cite{louseau} or as free-standing structures
\cite{bourgeois}.

%hc One among the many potential applications 
%hc of carbon nanotubes and nanocones is
%hc their use as cold electron sources in field emission applications
%hc \cite{ugarte}.
%hc In this context, chemical and thermal degradation are
%hc important issues. 
An application of carbon nanotubes and nanocones is
their use as cold electron sources in field emission displays
\cite{ugarte}, but
current-induced structural changes \cite{bonard} and
oxidation \cite{dean} have been found to deteriorate the 
devices.
%hc field emission 
%hc properties of carbon nanotubes. 
BN nanotubes, on the other hand, may prove more
thermally and chemically stable, as reported in a recent study
\cite{golberg}. Moreover, recent experiments have found that,
despite being insulating, BN nanotubes yield field emission 
currents comparable to those of carbon nanotubes 
at relatively low voltages \cite{zettl}. Besides the
%hc role played by geometrical field-enhancing factors in both carbon and
%hc BN nanotubes, understanding the field emission properties of BN
role played by geometrical field-enhancing factors,
understanding the field emission properties of BN
nanostructures
requires a study of their electronic structure, in particular 
at the tip region where conical structures are present \cite{foot}.
%hc the edge states which are determinant of the work function at the
%hc tip. In BN, the presence of two atomic species leads to a greater
%hc variety of distinct topological defects, when compared with
%hc carbon. Conical surfaces with four- or five-membered rings at the apex
%hc are examples of possible structures. In some cases, the 
%hc conical geometry implies the existence of antiphase-boundaries 
%hc (which are line defects in the case of two-dimensional systems),
%hc consisting of boron-boron (B-B) or nitrogen-nitrogen (N-N)
%hc bonds along the boundary. 
In the present study, we address the
electronic structure of several such conical BN structures  by
first-principles calculations. As discussed in detail below, our main
conclusion is that BN structures containing antiphase boundaries
depict non-bulk electron states that 
reduce the work function, relative to the bulk BN value, by as much as 2
eV. This could strongly increase the performance of BN field emission
devices relative to carbon analogues. 

A conical BN structure can be geometrically constructed 
by a ``cut and glue" process.
%hc characteristic of the formation of topological defects. 
%hc For instance,
%hc to create a single five-membered ring in a finite BN sheet, a sector
%hc of angle $\frac{2\pi}{6}$ is cut out starting from the center of any
%hc hexagon (so that one of the edges of the hexagon is removed) and the
%hc loose ends are joined together. The result is a conical structure with
%hc a pentagonal ring at its (truncated) apex.  The removal of a sector of
%hc angle $\frac{4\pi}{6}$ results in a square at the apex. Conversely, a
%hc seven-membered ring may be formed by inserting a wedge of angle
%hc $\frac{2\pi}{6}$, adding an extra edge to one of the hexagons. In this
%hc case, the result is a saddle-like structure with a heptagonal ring at
%hc the center.  Generally, removal of $n \frac{2\pi}{6}$ sectors (with
%hc $1\le n \le 5$) leads to a conical structure with a disclination of $n
%hc \times 60^\circ$ and a cone angle $\theta$ given by $sin(\theta/2) = 1
%hc - (n/6)$. 
Examples of structures built in this way and used in our calculations
are shown in Fig. 1. Fig. 1(a) shows a nanocone with a four-membered
ring at the apex, with a disclination angle of $2\times 60^\circ=120^\circ$.
The high strain at the tip of the structure is energetically
compensated by the absence of homopolar B-B or N-N bonds. This effect
is responsible for the high stability of BN fullerenes composed of
only four- and six-membered rings \cite{simone1,simone2}.  On the
other hand, rings with an odd number of atoms introduce non-BN bonds.
Figure 1(b) shows a structure which has two adjacent pentagons at the
apex and a single wrong bond. If a single pentagon is located at the
apex, non-BN bonds are formed along the cut, generating either a B-B
or an N-N antiphase-boundary. Examples of such antiphase boundaries are
indicated by arrows in Fig.s 1(c) and 1(d). The antiphase boundaries
can be of two kinds: a sequence of parallel B-B (Fig. 1d) or N-N
bonds, hereafter denoted molecular B and molecular N, respectively
(mol-B and mol-N, for simplicity); and a sequence of zig-zag B-B
(Fig. 1d) or N-N bonds, which we shall call zig-B and zig-N,
respectively.  Nanocones with large disclination angles (240$^\circ$
and 300$^\circ$) have also been observed \cite{bourgeois}.  Figure
1(e) shows one with two adjacent four-membered rings at the apex,
corresponding to a 240$^\circ$ disclination.  A saddle-like geometry
associated with a -60 $^\circ$ disclination, with 196 atoms, is shown
in Fig. 1(f),
%hc . An N-N bond is present at the central heptagon and the
%hc associated zig-N antiphase boundary is indicated in the figure by an
with a zig-N antiphase boundary indicated by an
arrow.

Each of these defects has a specific signature in the electronic
structure of the material. To address this issue, we apply first
principles calculations based on the Density Functional Theory
\cite{kohn} as implemented in the SIESTA program \cite{siesta1}. We
make use of norm-conserving Troullier Martins pseudopotentials \cite
{troullier} in the Kleinman-Bylander factorized form \cite {bylander},
and a double-$\zeta$ basis set composed of numerical atomic orbitals
of finite range.  Polarization orbitals are included for both,
nitrogen and boron atoms, and we use the generalized gradient
approximation (GGA) \cite {gga} for the exchange-correlation
potential.

The electronic structures of finite-size cones with different angles of
disclination is shown in Fig. 2. The solid lines
%hc correspond to the top of the valence band and the bottom of the conduction
correspond to the calculated valence and conduction band edges,$E_v$ and $E_c$,
%hc band 
of the isolated BN planar sheet. 
%hc The energy gap is 4.6 eV at the GGA level.
The small lines indicate the position of the electron states in the
region of the energy gap, relative to $E_v$, induced by the
topological transformation of the BN sheet. No alignment between bulk and
cone eigenvalues was necessary; $E_v$ coincides with the valence band
maximum of the cones within 0.1 eV, with the exception of the 300$^\circ$
disclination, for which there is a shift of 0.2 eV. We also performed additional
calculations for a 138-atom cone with a disclination of 300$^\circ$ to check
for cluster-size convergence. The resulting shift in the energies of the
gap levels was less than 0.06 eV.  

The first four columns in
Fig. 2 correspond to the four possible antiphase boundaries 
in a structure with
a pentagonal ring at the apex (60$^\circ$ disclination). The general
trend is that the presence of N-N bonds introduces occupied  states 
in the lower half of the band gap, 
while the states associated with B-B bonds are unoccupied and in the
upper half of the band gap. This is clearly seen in
Fig. 2 for the zig-N, mol-B and mol-N structures.  The exception to 
this rule is the zig-zag antiphase boundary of boron. In this case the defect
states are scattered over the entire range of the gap. This exception may be
partly caused by the geometry of this structure, shown in Fig.
1(c). Since a typical B-B bond length is greater than the B-N bond distance, 
the zig-zag antiphase boundary of boron is responsible for a large local
distortion in the structure, which should affect the electronic states.
Indeed, B-B bond distances of 1.7 \AA\ and B-B-B angles as small
as 70$^\circ$ are found in the antiphase boundary region. These values are to be
compared with the B-N bond length and B-N-B angle in the graphene
hexagonal network (1.45 \AA\  and 120$^\circ$, respectively).  In
molecular antiphase boundary nanocones, the B-B bonds are not adjacent, and as
a consequence, there is enough room to accomodate them without a great
distortion of the whole structure.

An important feature shown in Fig. 2 is the significant increase (by as
much as 2 eV) in the energy of the highest occupied electron state for 
all the BN structures that contain antiphase boundaries, with the
exception of the mol-B boundary. This would lead to a significant
reduction of the work function of BN nanocones and
cone-capped  nanotubes that contain antiphase boundaries. In
contrast, the structures without antiphase boundaries presented work
functions very similar to that of the planar BN sheet. These results
indicate that the performance of BN nanocones or cone-capped 
nanotubes in field emission devices would be very sensitive to the 
topology of the cones: those with antiphase boundaries would be
much better electron emitters than the ones without these line defects.

The relationship between non-BN bonds and gap states can be further
understood through the band structures of Fig. 3. In Fig. 3(a), we show the
electronic structure of an infinite boron nitride plane with 
boron and nitrogen zig-zag antiphase boundaries periodically separated
by $\sim$ 8.9 \AA. In Fig. 3(b), the same is shown
for a pair of molecular antiphase boundaries. The $k$-direction is along the
antiphase boundary. In both cases, the B-B and N-N bonds result in
additional bands close to the
conduction and valence band edges, respectively.  The result is the
closure of the gap for the zig-zag antiphase boundary and the narrowing 
of the gap for the molecular antiphase boundary.  
Note that the infinite-length antiphase boundaries lead to electron
bands (due to the the crystal periodicity along the direction of the
antiphase boundary) while the antiphase boundaries of the finite-size 
nanocones of the preceeding paragraphs lead to discrete states. The
former represent the electron states far from the cone tip, while the
latter include the states localized at the tip. However, in both cases 
the introduction of electron states associated
with the antiphase boundaries, covering partially or completely the range of the
electronic gap, may be associated with the surprisingly high values of
field emission currents at low voltages obtained in recent experiments
\cite{zettl}, since they may lead to a reduction of the work function
at the tip.

%hc Non-BN bonds are responsible for gap states even in the absence of an
%hc antiphase boundary. This can be seen from the 
%hc electronic structure calculations
%hc for a nanocone with two adjacent pentagons at the apex connected by a
%hc B-B bond, as shown in Fig. 1(b), and also for the similar structure
%hc with a N-N bond. The results, not shown in Fig. 2, follow the rule
%hc stated above, that is, an unoccupied gap state appears close to the
%hc conduction band edge for the B-B bond, and an occupied state is
%hc observed close to the valence band edge for the N-N bond.

The fifth column of Fig. 2 shows the change in the electronic structure 
due to the
topological operation that transforms a plane into a saddle-like
geometry (which has a disclination angle of -60$^\circ$). 
The result is nearly indistinguishable from the electronic
structure of the nanocone with one pentagon at the edge and the same
kind (zig-N) of antiphase boundary. This result indicates that the type of
chemical bonds present in the structure is more important than the
specific topology in determining the existence and position of
resonance and gap states, as long as the distortion is not as drastic
as in the case of the zig-zag antiphase boundary of boron.

The nanocone of Fig. 1(a), with a small disclination angles and 
no ``wrong'' bonds, does not depict gap states. In contrast, the
nanocone of Fig.1(e), with a large (240$^\circ$) disclination angle,
does have states in the gap, close to the conduction band, as indicated 
in the 6th column of Fig. 2. In the 7th column of the same figure
we show the electronic structure of a nanocone with a disclination of
300$^\circ$ and a molecular antiphase boundary of nitrogen. 
In this case, the deviation from the planar topology is so
large that several states are found in the gap, close to the
conduction and valence bands. These results indicate that extreme distortions
may induce gap states in addition to the antiphase-boundary states.
In that respect, we shall mention that nanocones with large disclination
angles (240$^\circ$ and 300$^\circ$) have been recently produced
\cite{bourgeois}.

%hc The last two columns of Fig. 2 show the electronic structure 
%hc of cones with the largest possible disclination angles (240$^\circ$ and 
%hc 300$^\circ$).
%hc They correspond to the structures that were observed in the
%hc experiments of Bourgeois et al \cite{bourgeois}, with disclinations of
%hc 240$^\circ$, as shown in Fig. 1(e), and 300$^\circ$. Despite having
%hc only B-N bonds, the 240$^\circ$ structure does have states in the gap,
%hc close to the conduction band.  For the disclination angle of
%hc 300$^\circ$ we considered a geometry with a molecular antiphase boundary of
%hc nitrogen. In this case, the deviation from the planar topology is so
%hc large that several states are found in the gap, close to the
%hcconduction and valence bands. These results indicate that extreme distortions
%hc may induce gap states in addition to the antiphase-boundary states.
%hc Another structure we investigated 
%hc contains a square at the apex.  As can be seen in Fig. 1(a), only
%hc B-N bonds are present and the disclination angle of 120$^\circ$ is not
%hc enough to introduce a significant distortion to the structure. Concerning
%hc the electronic structure, no gap states are found. Instead, resonance
%hc states appear close to the edge of the valence band.

In summary, we have investigated the electronic structure of boron
nitride nanocones.  We show that cones with
antiphase boundaries formed of B-B or N-N bonds introduce electron states 
in the energy-gap region, whereas cones without antiphase boundaries 
display resonant states near the band edges, related to the topological 
defect at the tip. Generally, states associated with the B-B and 
N-N bonds appear close to the
conduction and valence band edges, respectively. Gap states can also
be introduced by larger geometric distortions, such as those associated with
large disclinations angles.
Our results suggest that the changes in the electronic
structure of the cones containing antiphase boundaries 
should strongly reduce the work function of these BN nanostructures,
with important implications for understanding  the 
field emission properties of these systems.

The authors acknowledge support from CNPq, FAPEMIG and Instituto do Mil\^enio
em  Nanoci\^encias-MCT.

\newpage

\noindent
FIGURE CAPTIONS:

\vspace{0.5cm}
\noindent
Fig.1 - Some of the structures considered in the present study. 
B and N 
atoms are in grey and black, respectively, and hydrogen atoms saturate the
dangling bonds at the edges. (a) and (b) show cones with a disclination
angle $\theta=120^\circ$. (c) and (d) show cones with $\theta=60^\circ$. 
(e) is a cone with $\theta=240^\circ$, and (f)
is a saddle-like structure with $\theta=-60^\circ$. Some of the structures
contain non-BN bonds: (b) has a single B-B bond at the tip, (c) has a
zig-B antiphase boundary (see text), (d) has a mol-B antiphase boundary,
and (f) has a zig-N antiphase boundary. 
The arrows in (c), (d) and (f) indicate the antiphase boudaries.

\vspace{0.5cm}
\noindent
Fig. 2 - Electronic states of systems with different disclination angles  
$\theta$. The first four columns correspond to structures having 
$\theta=60^\circ$. The antiphase boundaries are Zig-N, Zig-B, Mol-B and 
Mol-N, respectively. 
The fifth
%hc column shows the gap states of the saddle-like geometry ($\theta=-60^\circ$
%hc with a nitrogen zig-zag antiphase boundary). The last two columns
column shows the gap states of the saddle-like structure of Fig. 1(f).
The last two columns
correspond to cones with large disclination angles,  
$\theta=240^\circ$ 
and $\theta=300^\circ$.
The electronic structure of the bulk BN is used as a reference with
the top of the valence band at the origin. 

\vspace{0.5cm}
\noindent
Fig. 3 - Electronic structure of infinite boron nitride planes, with
antiphase boundaries periodically separated by 8.9 \AA. 
The antiphase boundaries are zig-N and zig-B in (a) and  mol-N and mol-B in
(b).
The k-direction is along the antiphase boundaries in both cases. 

%\newpage


\begin{thebibliography}{10}
\bibitem{kroto} 
H. W. Kroto, J. R. Heath, S. C. O`Brien, R. F Curl, and
R. E. Smalley, {\it Nature}. {\bf 318}, 162 (1985).
\bibitem{iijima1} 
S. Iijima, {\it Nature}. {\bf 354}, 56 (1991).
\bibitem{hamada} 
N. Hamada, S. Sawada and A. Oshiyama. {\it Phys. Rev. Lett.}
{\bf 8}, 1579 (1992).
\bibitem{iijima2}
S. Iijima, T. Ichihashi, and Y. Ando. {\it Nature} {\bf 356}. 776 (1992).
\bibitem{charlier}
J. C. Charlier and G. M. Rignanese. {\it Phys. Rev. Lett.} {\bf 86}. 5970 (2001).
\bibitem{louseau}
A. Loiseau, F. Willaime, N. Demoncy, G. Hug, 
and H. Pascard {\it Phys. Rev. Lett.} 
{\bf 76}. 4737 (1996).
\bibitem{bourgeois} 
L. Bourgeois, Y. Bando, W. Q. Han and T. Sato. {\it Phys. Rev. B} {\bf 61}, 7686 (2000).
\bibitem{ugarte}
W. A. DeHeer, A. Chatelain, D. Ugarte, Science {\bf 270}, 1179 (1995).
\bibitem{bonard} 
J.-M. Bonard, J.-P. Salvetat, T. St\"ockli, W. A. de Heer, L. Forr\'o, and
A. Ch\^atelain, Appl. Phys. Lett., {\bf 73}, 918 (1998).
\bibitem{dean} K. A. Dean and B. R. Chalamala, Appl. Phys. Lett., {\bf 75}, 3017 (1999).
\bibitem{baylor} 
L. R. Baylor, V. I. Merkulov, E. D. Ellis, M. A. Guillorn, D. H. Lowndes,
A. V. Melechko, M. L. Simpson, and J. H. Whealton. {\it J. of Appl. Phys.} {\bf 91}, 4602
(2002). 
\bibitem{golberg} D. Golberg, Y. Bando, K. Kurashima, and T. Sato,
Scr. Mater, {\bf 44}, 1561 (2001).
\bibitem{zettl} J. Cumings and A. Zettl, {\it Electronic Properties of
Molecular Nanostructures}, edited by H. Kuzmany (AIP,
Melville, New York), p. 577.
\bibitem{foot}
This does not exclude the possibility of field emission originating from electron
states at edges of BN layers.
\bibitem{simone1} 
S. S. Alexandre, M. S. C. Mazzoni and H. Chacham. {\it Appl. 
Phys. Lett.} {\bf 75}, 61 (1999).
\bibitem{simone2} 
S. S. Alexandre, H. Chacham, R. W. Nunes {\it Phys. Rev. B} 
 {\bf 63}, 045402 (2001).
\bibitem{kohn}
W. Kohn, L. J. Sham. {\it Phys. Rev.} {\bf 140}, A1133 (1965).
\bibitem{siesta1}
D. Sanchez-Portal, P. Ordejon, E. Artacho and J. M. Soler. {\it Int. Journ. of 
Quant. Chem.} {\bf 65}, 453 (1997).
\bibitem{troullier} 
N. Troullier and J. L. Martins. {\it Phys. Rev. B.} {\bf 43}, 1993 (1991).
\bibitem{bylander}
L. Kleinman and D. M. Bylander. {\it Phys. Rev. Lett.} {\bf48}, 1425 (1982).
\bibitem{gga} 
J. P. Perdew, K. Burke, M. Ernzerhof. {\it Phys. Rev. Lett.} {\bf 77} 3865 (1996) 
\end{thebibliography}
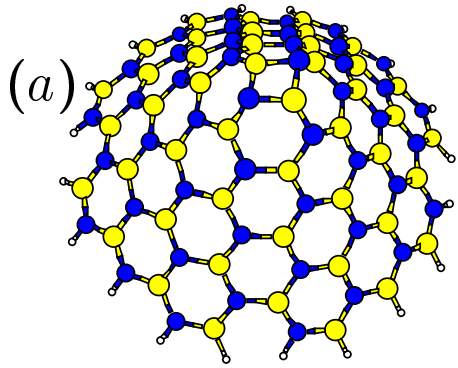
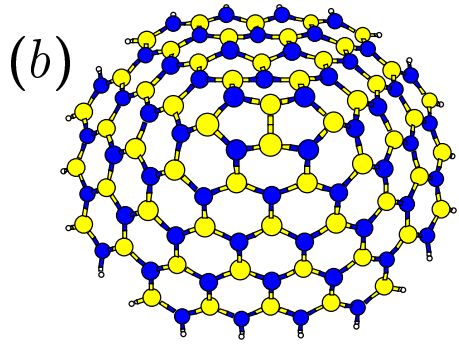
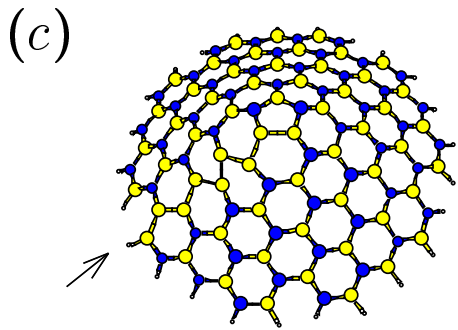
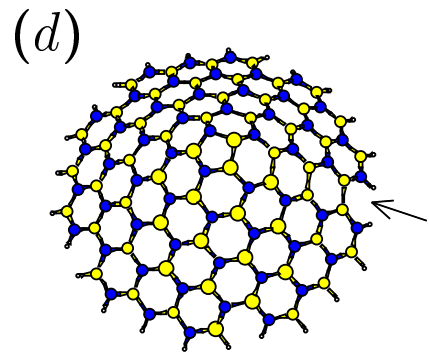
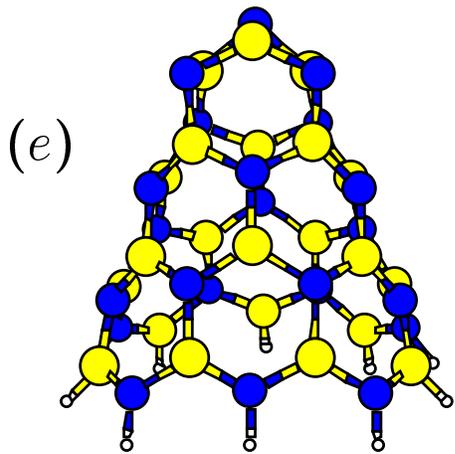
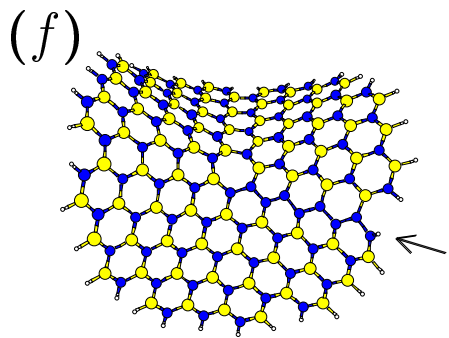
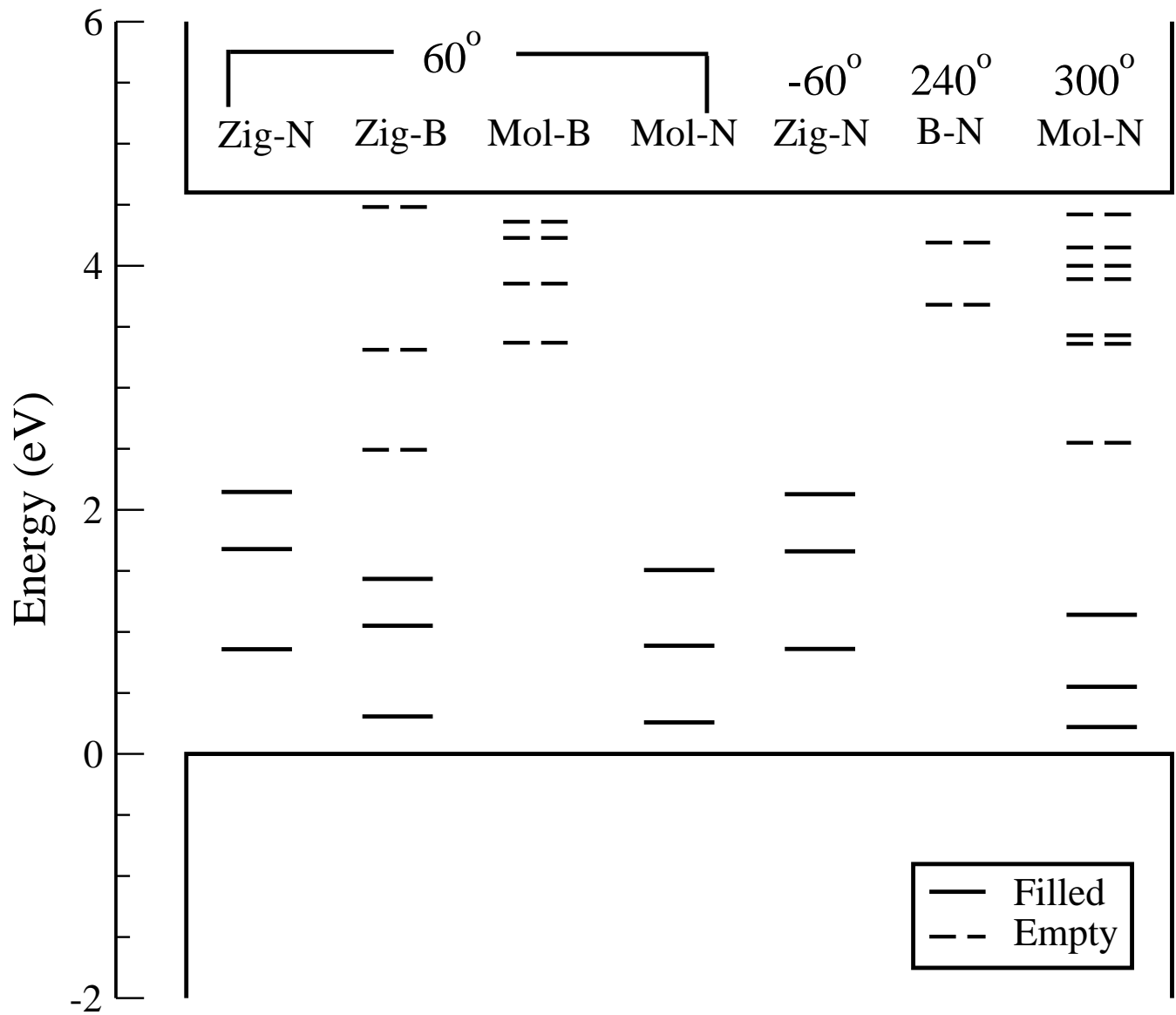
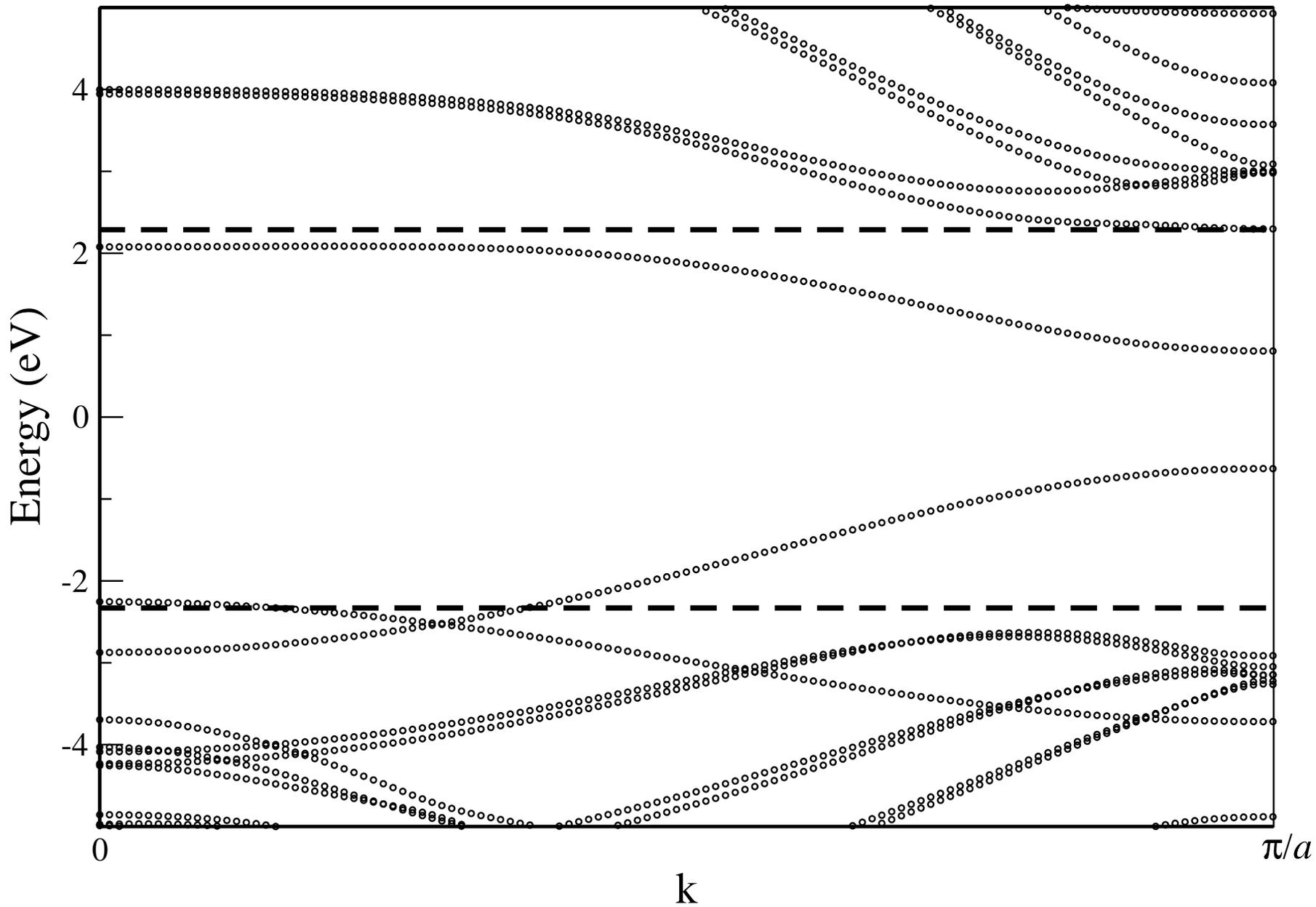
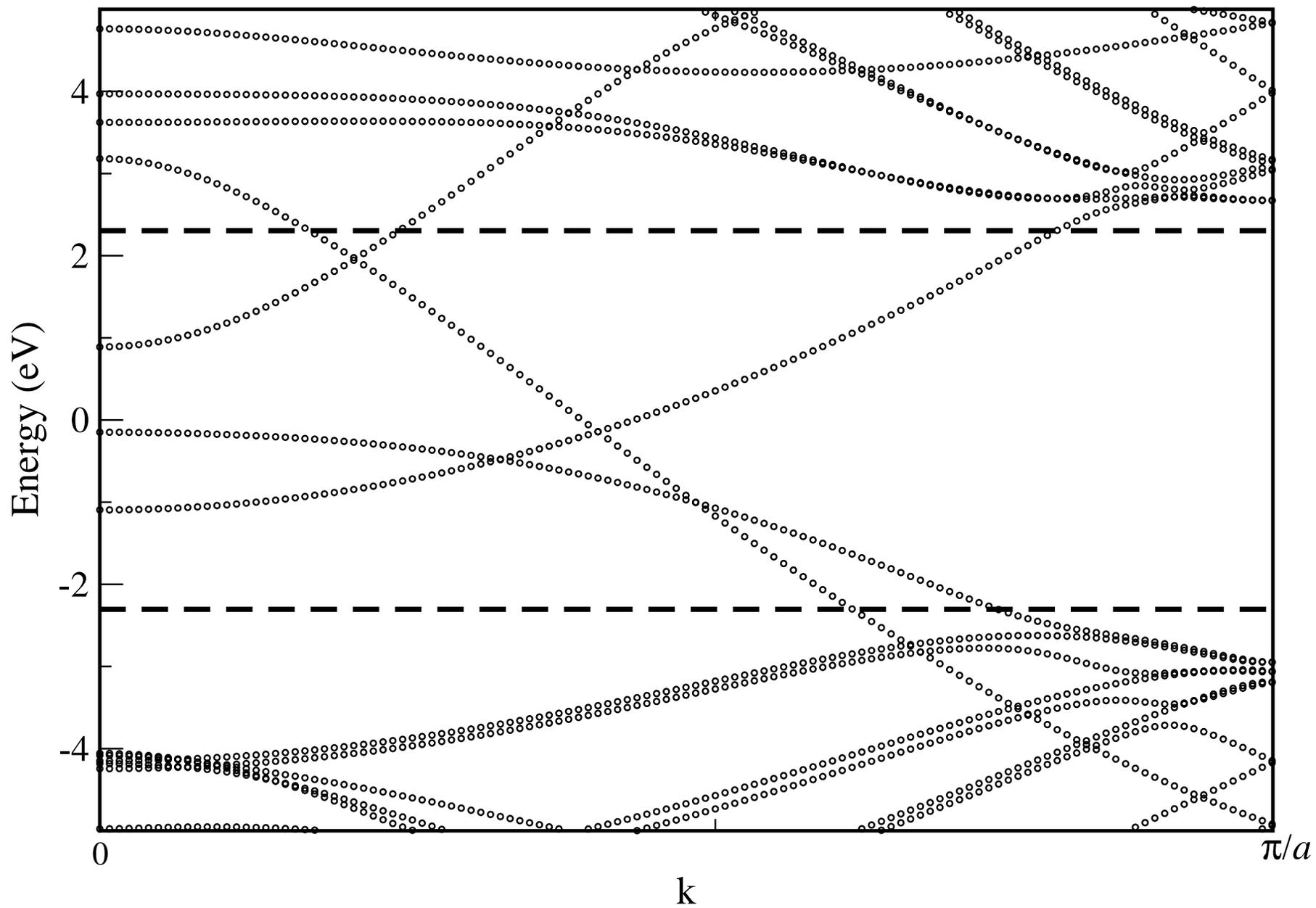
\end{document}